\documentclass[12pt,twoside]{article}

\begin{document}\hbadness=10000\thispagestyle{empty}
\pagestyle{myheadings}
\title{On quantum mechanics as constrained  
$N=2$ supersymmetric classical dynamics}
\author{$\ $\\
Hans-Thomas Elze 
\\ $\ $\\ 
Dipartimento di Fisica, Via F.\,Buonarroti 2, I-56127 Pisa, 
Italia\footnote{Permanent address: 
Instituto de F\'{\i}sica, Universidade Federal do Rio de Janeiro,  
C.P. 68.528, 21941-972 Rio de Janeiro, RJ, Brazil \,\,--\,\, E-mail: thomas@if.ufrj.br}
}
\vskip 0.5cm
\date{November 2004}
\maketitle
\vspace{-8.5cm}
\vspace*{8.0cm}
\begin{abstract}{\noindent
The Schr\"odinger equation is shown to be equivalent to a constrained Liouville equation 
under the assumption that phase space is extended to Grassmann algebra valued variables. 
For onedimensional systems, the underlying Hamiltonian dynamics has a $N=2$ supersymmetry. 
Potential applications to more realistic theories are briefly discussed.    
}\end{abstract}
\section{Introduction}
Since its very beginnings, there have been speculations on the possibility of deriving 
quantum theory from more fundamental dynamical structures, possibly deterministic 
ones. Famous is the discussion by Einstein, Podolsky and Rosen. This lead to the 
EPR paradox, which in turn was interpreted by its authors as indicating the need for a more complete fundamental 
theory \cite{EPR}. However, just as numerous have been attempts to prove no-go theorems prohibiting exactly such ``fundamentalism'', especially in local theories. 
This culminated in the studies of Bell, leading to the 
Bell inequalities \cite{Bell}. The paradox as well as the inequalities have come under experimental 
scrutiny in recent years. Here, and in general, no disagreement with quantum mechanics has been 
observed in the laboratory experiments on scales very large compared to the Planck scale.   
  
However, to this day, the feasible experiments cannot rule out the possibility that quantum mechanics emerges as an effective theory only on sufficiently large scales and can indeed be based on more fundamental models. 

Motivated by the unreconciled clash between general relativity and quantum theory, 't\,Hooft has 
argued in favour of model building in this context \cite{tHooft01} (see also further 
references therein). In various examples, the emergence of the usual Hilbert space structure and unitary 
evolution in deterministic classical models has been demonstrated 
in an appropriate large-scale limit. Particular emphasis has been placed on the 
observation that it is fairly simple to arrive at a Hilbert space formulation of the   
classical dynamics of systems with Hamiltonians which are linear in the momenta. However, 
at the same time, it is difficult to assure that the resulting 
emergent quantum models possess a well-defined groundstate, 
i.e., that their energy spectra are bounded from below. 

A new kind of gauge fixing or constraints implementing 
``information loss'' at a fundamental level have been invoked here. However, a unifying dynamical 
principle for the necessary truncation of the Hilbert space is still  
missing. Therefore, these models have to be constrained or discretized case by case \cite{tHooft01,ES02,Blasone04}.      

Various further arguments for deterministically induced quantum features have recently been proposed -- 
see, for example, the works collected in Part\,III of Ref.\,\cite{E04}, or Refs.\,\cite{ES02,Vitiello01,Smolin,Adler}, 
concerning discrete time models, statistical and/or dissipative systems,     
quantum gravity, and matrix models, among others.   

Most of these attempts to base quantum theory on a classical footing can be seen as 
variants of the earlier stochastic quantization procedures of Nelson \cite{Nelson} and of Parisi and Wu \cite{Parisi}, often accompanied by the problematic analytic continuation from 
imaginary (Euclidean) to real time (``Wick rotation''), in order to describe evolving 
systems. In distinction, 't\,Hooft's work points towards a truly dynamical understanding 
of the origin of quantum phenomena. 
   
In this note, my aim  is to report on a large class of 
deterministic classical systems where the {\it quantum mechanical features    
emerge from constrained classical dynamics}. In particular, 
based on the extension of classical phase space to variables which take their values in a Grassmann algebra, 
one obtains the Schr\"odinger equation with standard Hamiltonians. 
Extension of this work to interacting field theories is possible and will be 
considered elsewhere. The natural appearance of supersymmetry in this framework   
certainly deserves further study as well. 

Among the conceptual issues touched here, clearly 
the interpretation of the measurement 
process, of the ``collapse of the wave function'' in particular, must figure prominently, 
together with the quantum indeterminism and the wider philosophical implications of the 
algorithmic rules comprising quantum theory as a whole \cite{BZW}. It is left for future 
studies to find out, how the deterministic framework introduced here allows to see them 
in a new light. 

This letter is organized as follows. The (dis)similarity of the 
Liouville and the Schr\"odinger equations is demonstrated in Section\,2. In Section\,3,  
the discrepancy between both is overcome by introducing an extended phase space based 
on Grassmann algebra. This gives the Schr\"odinger equation a form which is suitable 
for reconstructing the underlying supersymmetric classical model in Section\,4. In the 
concluding section, I discuss some problems left to be clarified as well as interesting 
topics for further exploration.  

\section{(Dis)similarity of Liouville and Schr\"odinger equations}   
It will be convenient for the present argument to recall the Hilbert space formulation 
of classical statistical mechanics developed by Koopman and von\,Neumann \cite{KN}. 

Beginning with a $(2n)$-dimensional classical phase space ${\cal M}$, the coordinates 
are collectively denoted by $\varphi^a\equiv (q^1,\dots ,q^n;p^1,\dots ,p^n),\;a=1,\dots ,2n$, where 
$q,p$ stand for the usual coordinates and conjugate momenta. Given the time independent 
Hamiltonian $H(\varphi )$, the equations of motion are: 
\begin{equation}\label{eom}
\frac{\partial}{\partial t}\varphi^a=\omega^{ab}\frac{\partial}{\partial\varphi^b}H(\varphi ) 
\;\;, \end{equation} 
where $\omega^{ab}$ is the standard symplectic matrix, with a summation over indices appearing twice.      

Considering an ensemble of initial conditions, the evolution of the corresponding phase 
space density $\rho$ of a conservative system, which follows from Eqs.\,(\ref{eom}), is 
described by the Liouville equation:
\begin{equation}\label{Liouville} 
0=i\frac{\mbox{d}}{\mbox{d}t}\rho =i\partial_t\rho -\widehat{\cal L}\rho 
\;\;, \end{equation} 
where a convenient overall factor $i$ has been introduced, and the Liouville operator is: 
\begin{equation}\label{Liouville1}   
-\widehat{\cal L}\equiv\partial_pH\cdot i\partial_q-\partial_qH\cdot i\partial_p 
\;\;, \end{equation} 
in terms of partial derivatives with respect to phase space coordinates.  
 
In order to reformulate standard statistical mechanics in Hilbert space, 
the following two postulates are put forth: I) the density  
can be factorized as $\rho\equiv\Psi^*\Psi$, II) the complex valued amplitude or 
``state'' function $\Psi\in L^2$ itself obeys the Liouville equation (\ref{Liouville}). 

Furthermore, with the inner product defined by 
$\langle\Psi |\Phi\rangle\equiv\int\mbox{d}^nq\mbox{d}^np\;\Psi^*\Phi$, the Liouville operator  
is hermitean and the overlap $\langle\Psi |\Psi\rangle$ is a conserved quantity. Then, the 
Liouville equation also applies to $\rho =|\Psi |^2$, due to its linearity, and $\rho$ is 
consistently interpreted as probability density \cite{KN}.  

These results certainly remind one of the usual quantum mechanical formalism. In order to expose 
more clearly the similarity as well as a crucial difference, two further transformations of the  
Liouville equation for the state function $\Psi$ are useful.   

First of all, Fourier transformation replaces the momenta $p$ by new coordinates $Q$, 
$\Psi (q,p;t)=\int\mbox{d}^nQ\;\Psi (q,Q;t)\exp (-iQp)$, 
which yields: 
\begin{equation}\label{Fourier} 
i\partial_t\Psi =\Big\{ (-i\partial_Q)\cdot (-i\partial_q)+V'(q)\cdot Q\Big\} \Psi 
\;\;, \end{equation} 
where $V'(x)\equiv (\mbox{d}/\mbox{d}x)V(x)$, and a Hamiltonian with quadratic kinetic 
term has been assumed, in order to be explicit. 

Secondly, motivated by the definition of the quantum mechanical 
Wigner function, the following coordinate transformation is performed: 
\begin{equation}\label{Wigner} 
\sigma\equiv (q+Q)/\sqrt 2\;\;,\;\;\;\delta\equiv (q-Q)/\sqrt 2
\;\;. \end{equation}  
Thus, one obtains: 
\begin{equation}\label{Schroedinger} 
i\partial_t\Psi =\widehat{\cal H}\Psi\equiv\Big\{ -\frac{1}{2}(\partial_\sigma^{\;2}-\partial_\delta^{\;2})
+V'(\textstyle{\frac{\sigma +\delta}{\sqrt 2}})
\cdot\textstyle{\frac{\sigma -\delta}{\sqrt 2}}\Big\} \Psi
\;\;. \end{equation} 
This equation seems as close as one can get in a few steps from the 
classical Liouville equation to the Schr\"odinger equation. 

However, besides the characteristic doubling of the number of degrees of freedom, and their coupling 
in a particular form, there is a crucial difference between Eq.\,(\ref{Schroedinger}) 
and the Schr\"odinger equation. The spectrum of the effective Hamiltonian $\widehat{\cal H}$, 
generally, will not be bounded from below. This is related to the fact that 
$\widehat{\cal H}\rightarrow -\widehat{\cal H}$ under the interchange 
$\sigma\leftrightarrow\delta$.  

In any case, therefore, the 
above transformed classical theory, which is presented here in an appropriate Hilbert space form, 
lacks a stable groundstate and, therefore, does {\it not} qualify as a classical theory underlying 
quantum mechanics as an emergent description.  

This may suffice as a brief introduction of putting standard 
classical mechanics into Hilbert space form. Clearly, this is not limited to 
systems with a finite number of degrees of freedom.

\section{Extending phase space over Grassmann variables}   
It is obvious that the ``no-groundstate'' problem, which is encountered when trying to bridge 
the gap between Liouville and Schr\"odinger equations, requires a deep modification of the 
former, in order to be overcome.    
  
The following derivation will newly make use of ``pseudoclassical mechanics''.  
This notion has first been 
introduced in conjunction with the work of Casalbuoni and of Berezin and 
Marinov, who considered a {\it Grassmann variant of classical mechanics} in  
studying the classical dynamics of spin degrees of freedom as well as its      
quantized counterpart \cite{CB}. 

Classical mechanics based on Grassmann algebras has more recently 
found much attention,  
in order to illucidate the zerodimensional limit of classical and 
quantized supersymmetric field theories, see Refs.\,\cite{FDeW,MJ} 
and further references therein.  

{\it In all cases, so far, quantization is a second step, following a 
standard algorithm when applied to a given classical system. In distinction, the 
present work is concerned with the attempt to show that quantum mechanics  
emerges more directly from suitable classical structures without need for 
any of the known quantization procedures.} 

The considerations here will be based on the Grassmann algebra $\Lambda_2$. 
It is generated by two real odd (``fermionic'') elements $o_1,o_2$ obeying: 
\begin{equation}\label{odd} 
o_1^{\;2}=o_2^{\;2}=0=\{ o_1,o_2\} 
\;\;, \end{equation}
where the bracket denotes the {\it anticommutator}, $\{ A,B\}\equiv AB+BA$. 
In addition, there are two even (``bosonic'') elements $e_1,e_2$ which are given by: 
\begin{equation}\label{even} 
e_1\equiv 1\;\;,\;\;\;e_2\equiv o_1o_2 
\;\;. \end{equation} 
The even elements {\it commute} among themselves and with all other elements of the algebra. 
Furthermore, the definition of $e_2$ implies the nilpotency also of this element, 
$e_2^{\;2}=0$. -- 
The algebra $\Lambda_2$ is the simplest one which supports the concept of Fourier 
transformations with respect 
to even and odd supernumbers, to be employed in Section\,4.\footnote{Grassmann algebra  
and analysis over supernumbers are presented in detail by DeWitt \cite{FDeW}.}

The extension of the phase space pertaining to one classical degree of freedom is now 
introduced by the ``$\Lambda_2$-{\it postulate}'' that 
\begin{itemize} 
\item the variables $\sigma$ and $\delta$ take their values in the Grassmann algebra $\Lambda_2$ 
and are Grasssmann even and odd, respectively: 
\begin{equation}\label{extension} 
\sigma\equiv\sigma_ie_i\;\;,\;\;\;\delta\equiv\delta_io_i 
\;\;, \end{equation} 
where summation over $i=1,2$ is implied, and with $\sigma_i,\delta_i\in\mathbf{R}$.  
\end{itemize} 
Furthermore, the classical 
Liouville equation in the form of Eq.\,(\ref{Schroedinger}) is now replaced by:  
\begin{equation}\label{Schroedinger1} 
i\partial_t\Psi =\Big\{ -\frac{1}{2}(\partial_\sigma^{\;2}-\partial_\delta^{\;2})
+V'(\sigma +\delta)\cdot (\sigma -\delta )\Big\}_{\mbox{even}}\Psi
\;\;, \end{equation}  
where all terms are considered as Grassmann algebra valued; factors $1/\sqrt 2$ multiplying $V'$ 
and its argument in Eq.\,(\ref{Schroedinger}) have been absorbed conveniently into the 
definition of the potential. Furthermore, as indicated 
by the subscript $\{\dots\}_{\mbox{even}}$, only the Grassmann even part of the 
operator in brackets is taken here.  
 
These modifications of phase space and the  
evolution equation are related to {\it onedimensional quantum mechanics}, 
as will be shown next.\footnote{The analogous  
pointwise extension for field theories will be considered elsewhere.} 

In order to explore consequences of Eqs.\,(\ref{odd})--(\ref{Schroedinger1}), it is 
useful to expand 
the state function $\Psi$, incorporating the nilpotency of the odd Grassmann elements: 
$\Psi (\sigma ,\delta )\equiv\psi (\sigma )+\phi (\sigma )\delta$, where $\psi,\phi$ are 
Grassmann even yet possibly complex valued functions.   

Then, incorporating right derivatives, as discussed in Refs.\,\cite{FDeW}, it follows that 
$\partial_\sigma\Psi =\psi '+\phi '\delta$, $\partial_\delta\Psi =\phi$, 
$\partial_\sigma^{\;2}\Psi =\psi ''+\phi ''\delta$, $\partial_\delta^{\;2}\Psi =0$, 
and $\partial_\sigma\partial_\delta\Psi =\phi '=\partial_\delta\partial_\sigma\Psi$, 
where the primes denote ordinary derivatives, which are defined by the corresponding 
Taylor series, or similar, of the functions restricted to real arguments.   
Henceforth, all derivatives will be right derivatives, unless stated otherwise. 

Applying these derivatives and the expansion of the state 
function in Eq.\,(\ref{Schroedinger1}), the  
resulting equation can be decomposed with the help of the Grassmann algebra. Thus,   
one obtains two decoupled equations for the ``wave function'' $\psi$ and 
its ``shadow'' $\phi$: 
\begin{eqnarray}\label{wavefct} 
i\partial_t\psi (\sigma )&=&-\frac{1}{2}\psi ''(\sigma)+V'(\sigma )\sigma\psi (\sigma ) 
\;\;, \\ [1ex] \label{shadow} 
i\partial_t\phi (\sigma )&=&-\frac{1}{2}\phi ''(\sigma)+V'(\sigma )\sigma\phi (\sigma ) 
\;\;, \end{eqnarray}    
where it has also been assumed that $V(\sigma )$ is Grassmann even. 

Indeed, the {\it Schr\"odinger equation} and a formally identical {\it shadow equation} 
are obtained. They could naturally be combined into two-component form. 
This result followed by construction from the modification of  
the classical Liouville equation together with the extension of  
phase space over Grassmann variables. 

In itself, this may not be  
surprising. However, it will be demonstrated in Section\,4 that this theory  
presents a classical statistical mechanical description of a 
supersymmetric Hamiltonian system. Thus,  
quantum mechanics emerges from an underlying deterministic dynamics.  

Up to this point, the result is independent of the particular 
choice of $\Lambda_2$. Further decomposing both equations, making use of 
$\sigma =\sigma_ie_i=\sigma_1+\sigma_2e_2$, reproduces Eqs.\,(\ref{wavefct}),\,(\ref{shadow}) 
with $\sigma$ replaced by $\sigma_1$, its real ``body'' \cite{FDeW}, 
and yields additional higher-order derivative forms thereof, corresponding to applying 
$\partial_{\sigma_1}$ to both equations. The Schr\"odinger equation restricted to real 
variables, and correspondingly the usual quantum mechanical observables, are thus contained 
in the present framework.  

Several further remarks are in order here:  
\begin{itemize} 
\item There is no $\hbar$ in Eqs.\,(\ref{wavefct}),\,(\ref{shadow}) or, equivalently, units are 
such that $\hbar =1$. Thus, if introduced once and for all model potentials $V$ alike, it would 
merely act as a conversion factor of units. -- It is interesting to compare the present situation  
to the various points of view concerning the status of fundamental constants expressed  
in Ref.\,\cite{DOV}. 
\item The system has a stable groundstate, 
in particular for all potentials $V$, such that the onedimensional potential $xV'(x)$ 
yields bound states in quantum mechanics. 
\item There is no coupling between wave function and shadow. Such a coupling would be 
introduced, however, by a Grassmann odd contribution to the operator on the 
right-hand side of Eq.\,(\ref{Schroedinger1}).\footnote{It will be interesting to explore 
such a possibility with regard to the (breaking of) the supersymmetry of the underlying 
classical model (see Section\,4).} 
\end{itemize}
  
Furthermore, a probability amplitude interpretation of 
the state function, $\Psi (\sigma ,\delta )\equiv\psi (\sigma )+\phi (\sigma )\delta$, 
related to the wave function $\psi$ and its shadow $\phi$, 
is compatible with Eqs.\,(\ref{wavefct}),\,(\ref{shadow}) in the following sense. 
The normalization of $\Psi$ is time independent, 
\begin{equation}\label{norm}  
N\equiv\int\;\Psi^*\Psi\mbox{d}M\equiv\int\;\Psi^*\Psi (a+b\delta )\mbox{d}\sigma\mbox{d}\delta 
=Z\int\Big\{ b\psi^*\psi+a(\phi^*\psi +\phi\psi^*)\Big\}\mbox{d}\sigma\;=\;\mbox{const} 
\;\;, \end{equation} 
since the wave and shadow function can be expanded in terms of the same set of orthonormal 
stationary states. A general measure in terms of 
two constants $a,b\in\mathbf{C}$ has been assumed and the rules for integration 
over Grassmann odd variables have been applied: $\int\mbox{d}\delta=0$, $\int\delta\mbox{d}\delta=Z$, 
with $Z\in\mathbf{C}$ a conventional factor \cite{FDeW}. 

The normalization can be chosen real, with $0\leq N\leq 1$, for $\psi$ and $\phi$ properly normalized to 
unity, by setting $b=(2Z)^{-1}=2a$. For $\phi =\psi$, one has $N=1$, while in all other cases 
probability appears to be missing. This might have phenomenological implications, similar to the 
``negative probability'' considered by Feynman \cite{Feynman}. 
 
The (pseudo-)Liouville 
Eq.\,(\ref{Schroedinger1}) will be the starting point of the  
reconstruction of the classical mechanics which underlies the Schr\"odinger 
and shadow equations, which follows next.  

\section{Supersymmetric Hamiltonian dynamics beneath \\ Schr\"odinger and shadow equations}   
The strategy here is simple. As close as possible, the derivations of Section\,2, which led 
from the classical equations of motion (\ref{eom}) to the Liouville equation (\ref{Schroedinger}) 
in Hilbert space, will be reversed, duly taking the $\Lambda_2$-postulate (\ref{extension}) of Section\,3 
into account. In this way, a classical dynamics will be found for which the analogous 
Liouville equation is Eq.\,(\ref{Schroedinger1}), i.e.,   
is equivalent to the Schr\"odinger and shadow equations, 
Eqs.\,(\ref{wavefct}),\,(\ref{shadow}) respectively.  

To begin with, the coordinate transformation (\ref{Wigner}) is undone by introducing: 
\begin{equation}\label{WignerR} 
q\equiv (\sigma +\delta )/\sqrt 2\;\;,\;\;\;Q\equiv (\sigma -\delta )/\sqrt 2
\;\;, \end{equation}  
where $\sigma,\delta$ are the Grassmann even and odd variables defined in Eqs.\,(\ref{extension}). Note that $[q,Q]=0$.   
Similarly, the derivatives, 
$\partial_{q(Q)}\equiv(\partial_\sigma +(-)\partial_\delta )/\sqrt 2$, commute with each other.

Incorporating this transformation, the Liouville Eq.\,(\ref{Schroedinger1}) 
turns into the Grassmann analogue of Eq.\,(\ref{Fourier}), 
which retains its form. However, being defined as sum and difference of the same Grassmann 
even and odd terms in Eqs.\,(\ref{WignerR}), 
this specific structure of the variables $q,Q$ has to be enforced by {\it constraints}. 
They can be stated in different ways. 

A geometric set of 
constraints is: $(q-Q)^2=0$ (``distance zero''), $q-Q=q^*-Q^*$ (``reality''), 
\underline{and} $qQ=(q+Q)^2/4$ (``geometric mean squared = arithmetic mean squared''). 
Equivalently, one may demand $[q,Q]=0$ instead of the last requirement.  

More simply, however, one may require $q_e=Q_e$ and 
$q_o=-Q_o$, where the subscripts ``$e,o$'' refer to even and odd components of the respective variable. Thus, at the end of the present section,  
the constraints will be implemented by integrating the Liouville equation derived  
in the following, or, equivalently, the Grassmann analogue of Eq.\,(\ref{Fourier}),   
with Dirac $\delta$-function weights:  
$\int\;\dots\;\delta (q_e-Q_e)\delta (q_o+Q_o)\mbox{d}Q_e\mbox{d}Q_o\;$, where the order of 
Grassmann odd terms is important. The $\delta$-function (distribution) for Grassmann algebra valued variables forms the basis for the related generalized theory of Fourier transformation 
\cite{FDeW}. 

Next, in fact, the Fourier transformation preceding Eq.\,(\ref{Fourier}) will be undone. 
Properly defining the Fourier transformation over Grassmann variables needs some care and  
has been elaborated by DeWitt \cite{FDeW}. One has to 
proceed in two steps: 
\begin{eqnarray}
f(q,Q)\equiv f(q,Q_e,Q_o)&=&\int\;f(q,Q_e;p_o)\exp (ip_oQ_o)\frac{\mbox{d}p_o}{\sqrt{2\pi}} 
\nonumber \\ [1ex] \label{FourierG} 
&=&\int\;\exp (ip_eQ_e+ip_oQ_o)f(q;p_e,p_o)
\frac{\mbox{d}p_e}{2\pi}\frac{\mbox{d}p_o}{\sqrt{2\pi}}
\;\;. \end{eqnarray} 
Then, employing the appropriate partial integrations \cite{FDeW} where necessary, one calculates: 
\begin{eqnarray}
Q\Psi (q,Q)&\equiv&(Q_e+Q_o)\Psi (q,Q_e,Q_o)
\nonumber \\ [1ex] \label{QFourierG}
&=&\int\;\exp (ip_eQ_e+ip_oQ_o)(i\partial_{p_e}+i\partial_{p_o})\Psi (q;p_e,p_o)
\frac{\mbox{d}p_e}{2\pi}\frac{\mbox{d}p_o}{\sqrt{2\pi}} 
\\ [1ex]  
\partial_Q\Psi (q,Q)&\equiv&(\partial_{Q_e}+\partial_{Q_o})\Psi (q,Q_e,Q_o)
\nonumber \\ [1ex] \label{partFourierG}
&=&\int\;\exp (ip_eQ_e+ip_oQ_o)i(p_e-p_o)\Psi (q;p_e,p_o)
\frac{\mbox{d}p_e}{2\pi}\frac{\mbox{d}p_o}{\sqrt{2\pi}} 
\;\;. \end{eqnarray} 
Generally, the ordering of factors is important, due to Grassmann integration and algebra. 

Incorporating the sequence of transformations discussed in this section, so far, 
the {\it Liouville equation} (\ref{Schroedinger1}) attains the more familiar looking form: 
\begin{equation}\label{psLiouville} 
\partial_t\Psi =-\Big\{ 
p_e\partial_{q_e}-p_o\partial_{q_o}-V'(q_e)\partial_{p_e}-V''(q_e)q_o\partial_{p_o}
\Big\}\Psi 
\;\;, \end{equation} 
where $\Psi\equiv\Psi (q_e,q_o;p_e,p_o;t)$. 
Note that the operator on the right-hand side comprises only Grassmann even terms, as   
demanded before.  

It will now be shown that Eq.\,(\ref{psLiouville}) is, in fact, the Liouville 
equation for the Lagrangian: 
\begin{equation}\label{psLagrangian} 
L\equiv\dot q_e\dot q_o-V'(q_e)q_o  
\;\;, \end{equation} 
with $V'(q_e)$ Grassmann even.\footnote{This Lagrangian apparently has not been studied 
before, which might be related to the fact that the action obtained by integrating $L$ 
over real time is Grassmann odd. However, in line with the present attempt to find a   
classical structure {\it beneath} quantum mechanics, no (path integral) quantization of the 
model is intended.}    

Introducing the canonical momenta, 
\begin{equation}\label{momenta} 
P_{e,o}\equiv\frac{\partial L}{\partial\dot q_{e,o}}=\dot q_{o,e} 
\;\;, \end{equation} 
the Hamiltonian becomes: 
\begin{equation}\label{psHamiltonian} 
H\equiv P_e\dot q_e+P_o\dot q_o-L=P_eP_o+V'(q_e)q_o  
\;\;. \end{equation} 
In turn, this leads to the equations of motion:
\begin{eqnarray}\label{pseom1}
\dot q_e=\frac{\partial H}{\partial P_e}=P_o&,&\;\;\dot q_o=\frac{\partial H}{\partial P_o}=P_e
\;\;, \\ [1ex]\label{pseom2}
\dot P_e=-\frac{\partial H}{\partial q_e}=-V''(q_e)q_o&,&\;\;\dot P_o=-\frac{\partial H}{\partial q_o}=-V'(q_e) 
\;\;. \end{eqnarray}
or, in second-order form:  
\begin{equation}\label{pseoms} 
\ddot q_e=-V'(q_e)\;\;,\;\;\;\ddot q_o=-V''(q_e)q_o 
\;\;. \end{equation}
These equations imply that the Hamiltonian is a constant of motion, as expected.

Furthermore, identifying $p_e\equiv P_o$ and $-p_o\equiv P_e$, 
consistently with the Grassmann nature of each variable, it is now seen 
that the Liouville equation (\ref{psLiouville}) is indeed 
of the usual form: 
\begin{equation}\label{psLiouville1} 
\partial_t\Psi =-\Big\{\frac{\partial H}{\partial P_e}\partial_{q_e}+\frac{\partial H}{\partial P_o}\partial_{q_o}
-\frac{\partial H}{\partial q_e}\partial_{P_e}-\frac{\partial H}{\partial q_o}\partial_{P_o}\Big\}\Psi 
=-\{ H,\Psi\}_{\mbox{PB}}
\;\;, \end{equation}  
cf. Section\,2. Here the graded antisymmetric Poisson bracket is introduced \cite{FDeW}. 
For any pair of dynamical variables $A,B$ it is defined by: 
\begin{equation}\label{PB} 
\{ A,B\}_{\mbox{PB}}\equiv A\{ 
\stackrel{\leftharpoonup}{\partial_{P_e}}\stackrel{\rightharpoonup}{\partial_{q_e}}
+\stackrel{\leftharpoonup}{\partial_{P_o}}\stackrel{\rightharpoonup}{\partial_{q_o}}
-\stackrel{\leftharpoonup}{\partial_{q_e}}\stackrel{\rightharpoonup}{\partial_{P_e}}
-\stackrel{\leftharpoonup}{\partial_{q_o}}\stackrel{\rightharpoonup}{\partial_{P_o}}
\} B
\;\;, \end{equation} 
where left and right derivatives are indicated explicitly. 

The action associated with the above Lagrangian and the equations of motion have various interesting symmetry properties.

Using the decomposition $q_e(t)=q_1(t)+q_2(t)e_2$, together with the defining 
properties (\ref{odd}) and (\ref{even}) of $\Lambda_2$, one finds that $L$ does not contain 
derivatives of $q_2$. Therefore, it presents a real parameter, which can be eliminated 
by a translation.\footnote{Replacing the Grassmann algebra $\Lambda_2$ by $\Lambda_3$, 
for example, all coordinates would be dynamical. However, elimination 
of $q_2$ and, thus, of $e_2$ may be wellcome, since $e_2$ is imaginary \cite{FDeW}. This 
renders $L,H$, etc. real.} 
Thus, the first of Eqs.\,(\ref{pseoms}) becomes an ordinary 
equation of motion, while the second equation describes a parametrically coupled ``fermionic'' 
oscillator. 

For the harmonic oscillator, with $V'(q_e)=\bar Vq_e$, and for a constant potential
the Eqs.\,(\ref{pseoms}) decouple and are supersymmetric under the discrete interchange  
$q_e\leftrightarrow q_o$. 

This suggests to look for supersymmetry also in the case of an 
arbitrary potential. 
Indeed, the system described by the Lagrangian $L$ of Eq.\,(\ref{psLagrangian}) is 
invariant under the transformation: 
\begin{equation}\label{trans1}
q_e\;\longrightarrow\;q_e+\epsilon q_o 
\;\;, \end{equation} 
where $\epsilon$ is a Grassmann even infinitesimal parameter. To this is associated a conserved even  
``charge'' $C$, which is obtained by the usual Noether method:  
\begin{equation}\label{C} 
C=q_o\dot q_o=q_oP_e 
\;\;. \end{equation}  
Similarly, there is a second transformation which leaves the dynamics invariant: 
\begin{equation}\label{trans2}  
q_o\;\longrightarrow\;q_o+\epsilon\dot q_e 
\;\;, \end{equation} 
with associated Noether charge: 
\begin{equation}\label{He} 
H_e=\frac{1}{2}\dot q_e^{\;2}+V(q_e)=\frac{1}{2}P_o^{\;2}+V(q_e)   
\;\;, \end{equation} 
i.e., the energy of the even degree of freedom, particularly when $q_e=q_1\in\mathbf{R}$, 
as discussed. 

Summarizing the symmetry properties, the above constants of motion satisfy the following 
$N=2$ {\it supersymmetry algebra} generated by the two charges: 
\begin{equation}\label{susy} 
\{ C,H_e \}_{\mbox{PB}}=H 
\;\;, \end{equation}  
with the Poisson bracket of Eq.\,(\ref{PB}). Furthermore,    
for all combinations of $A,B\in\{ C,H_e,H\}$ other than the above, 
one finds $\{ A,B\}_{\mbox{PB}}=0$.  

At last, the constraints still should be implemented on the Liouville equation 
(\ref{psLiouville}), or Eq.\,(\ref{psLiouville1}), in more compact form. As mentioned 
before, this is achieved by integrating this equation 
with suitable $\delta$-function weights: 
\begin{eqnarray}  
&\;&\int\Big (\int\exp (ip_eQ_e+ip_oQ_o)
\Big [\partial_t\Psi +\{ H,\Psi\}_{\mbox{PB}}\Big ]
\frac{\mbox{d}p_e}{2\pi}\frac{\mbox{d}p_o}{\sqrt{2\pi}}\Big ) 
\delta (q_e-Q_e)\delta (q_o+Q_o)\mbox{d}Q_e\mbox{d}Q_o
\nonumber \\ [1ex]\label{cpsLiouville1} 
&\;&=\;
\int\exp (ip_eq_e+ip_oq_o)
\Big [\partial_t\Psi +\{ H,\Psi\}_{\mbox{PB}}\Big ]
\frac{\mbox{d}p_e}{2\pi}\frac{\mbox{d}p_o}{\sqrt{2\pi}}\;=\;0
\;\;, \end{eqnarray}
where $p_e=P_o$ and $-p_o=P_e$, as before.
Note that the Fourier integrals were eliminated in the derivation 
of Eq.\,(\ref{psLiouville}) by the inverse transformation. Here, they have to be kept, 
since the dependence 
on the coordinates $Q_e,Q_o$ enters the integrations effecting the constraints.   

The equations (\ref{cpsLiouville1}) (remaining variables $q_e,q_o,t$)  
present the main result of this section. While the left-hand side shows the implementation 
of the local constraints on the coordinates $q,Q$, the second expression involves     
an integro-differential operator, of course, due to Fourier transformation.  
The resulting equation is automatically solved by all solutions $\Psi$ of the classical 
Liouville equation. -- 
It is conceivable that further solutions exist corresponding 
to solutions of the Liouville equation with source terms $s$, such that 
$\int\exp (ip_eq_e+ip_oq_o)s(q_e,q_o;p_e,p_o;t)
\mbox{d}p_e\mbox{d}p_o=0$. 
Study of their existence, properties, and relevance is left as an important topic for 
future work. 
 
Summarizing, it has been shown here that 
the constrained (pseudo-)Liouville equation (\ref{cpsLiouville1}), pertaining 
to the deterministic 
supersymmetric Hamiltonian dynamics of the model defined by the Lagrangian 
(\ref{psLagrangian}), follows from  
the Schr\"odinger and shadow equations (\ref{wavefct}) and (\ref{shadow}). 
Inversely, the quantum mechanical equations follow from the classical 
Liouville equation (\ref{psLiouville}), or Eq.\,(\ref{psLiouville1}),  
again implementing the constraints at the end. 

\section{Conclusions} 
Presently, it has been shown how the Schr\"odinger and shadow equations emerge from 
underlying classical dynamics. This may certainly be questioned in many respects. 
It might violate one or the other of the assumptions of existing 
no-go theorems relating to hidden variables theories. 
However, it is unknown whether those assumptions will be relevant for a future 
fundamental theory of physics at the Planck scale. Therefore, it is a valid option to try  
and reconstruct quantum theory as an emergent or effective theory for presently 
accessible scales \cite{tHooft01}--\cite{Adler}. 

It seems interesting to further explore the demonstrated connection between quantum mechanical 
and deterministic classical structures which makes no use of any of the known 
quantization procedures. Here, instead, the quantum mechanical 
features arise in a constrained phase space description of dynamics which is based on the Grassmann 
algebra valued variables of an underlying classical supersymmetric model.   

The wave function and its shadow appear together in   
Eqs.\,(\ref{wavefct}) and (\ref{shadow}), respectively. Does the shadow contribute to 
observables? What is the interpretation of observables related to the ``soul'' \cite{FDeW}  
of the Grassmann algebra valued variables? 
Detailed solution of the dynamics of quantum mechanics 
textbook examples should and will be repeated elsewhere, in order to further illucidate 
the description  
by the pseudoclassical Liouville equation (\ref{psLiouville}), or Eq.\,(\ref{psLiouville1}), 
and especially by its constrained version, Eq.\,(\ref{cpsLiouville1}).  

The extension to higher-dimensional classical models or field theories seems 
straightforward. However, what are 
the physical implications of supersymmetry (or its breaking) of the underlying classical 
system, as seen in our formalism? 
Where do physical fermions come in?
Most interestingly, in emergent quantum field theories, 
is there a relation of their typical divergences to the necessary constraints on the 
Grassmann structure of the relevant phase space variables, cf. Eqs.\,(\ref{WignerR}) and 
the ensuing remarks? How do these constraints interfere with intrinsic constraints, 
for example, Gauss' law in gauge theories or M(atrix) theory \cite{Smolin,Adler,Taylor}?      

Clearly, there is room for further work. In the long run, this could lead to a reassessment of 
the fundamental role played by quantum theory in our description 
of natural phenomena. One may also ponder anew about the conceptual  
issues, briefly alluded to in the introduction, which surround quantum theory 
in its present form. 
 
\subsection*{Acknowledgements} 
I am grateful to A.\,DiGiacomo 
for kind hospitality and discussions at the Dipartimento di Fisica in Pisa and 
to R.\,D'Auria, C.\,Kiefer, and G.\,Vitiello for helpful remarks and discussions. 




\begin{thebibliography}{99}
\bibitem{EPR} A.\,Einstein, B.\,Podolsky and N.\,Rosen, Phys.\,Rev. {\bf 47}, 777 (1935).  
\bibitem{Bell} J.S.\,Bell, {\it ``Speakable and Unspeakable in Quantum Mechanics''} (Cambridge U.  
Press, Cambridge, 1987). 
\bibitem{tHooft01} G.\,'t\,Hooft, {\it Quantum Mechanics and Determinism}, in: Proceedings of the Eighth Int. Conf. 
on ``Particles, Strings and Cosmology'', ed. by P.\,Frampton and J.\,Ng (Rinton Press, Princeton, 2001), p.275; 
hep-th/0105105 ; \\ see also: {\it Determinism Beneath Quantum Mechanics}, quant-ph/0212095 .
\bibitem{ES02} H.-T.\,Elze and O.\,Schipper, Phys.\,Rev. {\bf D66}, 044020 (2002); \\  
H.-T.\,Elze, Phys.\,Lett. {\bf A310}, 110 (2003); Physica {\bf A344}, 478 (2004). 
\bibitem{Blasone04} M.\,Blasone, P.\,Jizba and H.\,Kleinert, 
{\it Path Integral Approach to 't\,Hooft's Derivation of Quantum from Classical Physics}, quant-ph/0409021 .
\bibitem{E04} {\it ``Decoherence and Entropy in Complex Systems''}, ed. by H.-T.\,Elze, 
Lecture Notes in Physics, Vol. 633 (Springer-Verlag, Berlin Heidelberg New York, 2004). 
\bibitem{Vitiello01} M.\,Blasone, P.\,Jizba and G.\,Vitiello, Phys.\,Lett. {\bf A287}, 205 (2001); \\  
M.\,Blasone, E.\,Celeghini, P.\,Jizba and G.\,Vitiello, Phys.\,Lett. {\bf A310}, 393 (2003).  
\bibitem{Smolin} L.\,Smolin, {\it Matrix Models as Non-Local Hidden Variables Theories}, hep-th/0201031 ; \\ 
F.\,Markopoulou and L.\,Smolin, {\it Quantum Theory from Quantum Gravity}, gr-qc/0311059 . 
\bibitem{Adler} S.L.\,Adler, {\it Statistical Dynamics of Global Unitary Invariant Matrix Models as 
Pre-Quantum Mechanics}, hep-th/0206120 . 
\bibitem{Nelson} E.\,Nelson, Phys.\,Rev. {\bf 150}, 1079 (1966). 
\bibitem{Parisi} G.\,Parisi and Y.S.\,Wu, Sci.\,Sin. {\bf 24}, 483 (1981); \\  
P.H.\,Damgaard and H.\,H\"uffel, Phys.\,Rep. {\bf 152}, 227 (1987). 
\bibitem{BZW} {\it ``Quantum Theory and Beyond''}, ed. by T.\,Bastin (Cambridge U. Press, 
Cambridge, 1971); \\    
{\it ``Quantum Theory and Measurement''}, ed. by J.A.\,Wheeler and W.H.\,Zurek (Princeton U.  Press, 
Princeton, 1980).
\bibitem{KN} B.O.\,Koopman, Proc.\,Nat.\,Acad.\,Sci. (USA) {\bf 17}, 315 (1931); \\ 
J.\,von\,Neumann, Ann. Math. {\bf 33}, 587 (1932); ibid. {\bf 33}, 789 (1932).   
\bibitem{CB} R.\,Casalbuoni, Nuovo\,Cim. {\bf 33A}, 389 (1976); \\  
F.A.\,Berezin and M.S.\,Marinov, Ann.\,Phys. (NY) {\bf 104}, 336 (1977). 
\bibitem{FDeW} P.G.O.\,Freund, {\it ``Introduction to Supersymmetry''} (Cambridge U.  
Press, Cambridge, 1986); \\ B.\,DeWitt, {\it ``Supermanifolds''}, 2nd ed. (Cambridge U.  
Press, Cambridge, 1992).  
\bibitem{MJ} N.S.\,Manton, J.\,Math.\,Phys. {\bf 40}, 736 (1999); \\ 
G.\,Junker, S.\,Matthiesen and A.\,Inomata, {\it Classical and quasi-classical aspects 
of supersymmetric quantum mechanics}, hep-th/95102230 .
\bibitem{DOV} M.J.\,Duff, L.B.\,Okun and G.\,Veneziano, JHEP {\bf 03}, 023 (2002). 
\bibitem{Feynman} R.P.\,Feynman, {\it Negative Probability}, in: {\it ``Quantum implications: essays in 
honour of David Bohm''}, ed. by B.J.\,Hiley and F.D.\,Peat (Routledge \& Kegan Paul, New York, 1987), p.235. 
\bibitem{Taylor} W.\,Taylor, Rev.\,Mod.\,Phys. {\bf 73}, 419 (2001). 
\end{thebibliography}
\end{document}